\documentclass[final,5p,times,twocolumn,sort&compress]{elsarticle}

\usepackage{epsfig}
\usepackage{amsmath}
\usepackage{url}
\usepackage[colorlinks,linkcolor=blue,anchorcolor=blue,citecolor=blue,urlcolor=blue]{hyperref}
\usepackage{graphicx}
\usepackage{lineno}
\usepackage{color}
\usepackage{ulem}
\usepackage{multirow}

\journal{Physics Letter B}
\usepackage{threeparttable}
\usepackage{multirow}
\usepackage{tablefootnote}
\usepackage{booktabs}
\usepackage{float}
\usepackage{latexsym}
\usepackage{amssymb}
\usepackage{lipsum}
\usepackage{latexsym}
\usepackage{amssymb}

\begin{document}

\begin{frontmatter}

\title{Single-proton removal reaction in the IQMD+GEMINI model benchmarked by elemental fragmentation cross sections of $^{29-33}\mathrm{Si}$ on carbon at $\sim$230~MeV/nucleon}
\author[Beihang]{Guang-Shuai~Li}

\author[Beihang]{Bao-Hua~Sun\corref{cor1}}
\ead{bhsun@buaa.edu.cn}

\author[Sino]{Jun~Su\corref{cor1}}
\ead{sujun3@mail.sysu.edu.cn}

\author[Beihang,RCNP]{Isao~Tanihata}

\author[IMP,Beihang]{Satoru~Terashima}

\author[Peking,GSI,Beihang]{Jian-Wei~Zhao}

\author[Sino]{Er-Xi~Xiao}

\author[Beihang]{Ji-Chao~Zhang}

\author[Beihang]{Liu-Chun~He}

\author[Beihang]{Ge~Guo}

\author[Sichuan]{Wei-Ping~Lin}

\author[Beihang]{Wen-Jian~Lin}

\author[Beihang]{Chuan-Ye~Liu}

\author[IMP]{Chen-Gui~Lu}

\author[Sino,IMP]{Bo~Mei}

\author[Beihang]{Dan-Yang~Pang}

\author[Beihang]{Ye-Lei~Sun}

\author[IMP]{Zhi-Yu~Sun}

\author[Beihang]{Meng~Wang}

\author[Beihang]{Feng~Wang}

\author[Beihang]{Jing~Wang}

\author[IMP]{Shi-Tao~Wang}

\author[Beihang]{Xiu-Lin~Wei}

\author[IMP,Beihang]{Xiao-Dong~Xu}

\author[Beihang]{Jun-Yao~Xu}

\author[Beihang]{Li-Hua~Zhu}

\author[IMP]{Yong~Zheng}

\author[Beihang]{Mei-Xue~Zhang}

\author[IMP]{Xue-Heng~Zhang}

\date{\today}

\cortext[cor1]{Corresponding Author}
\address[Beihang]{School of Physics, Beihang University, Beijing 100191, China}
\address[Sino]{Sino-French Institute of Nuclear Engineering and Technology, Sun Yat-sen University, Zhuhai 519082, China}
\address[IMP]{Institute of Modern Physics, Chinese Academy of Sciences, Lanzhou 730000, China}
\address[Peking]{State Key Laboratory of Nuclear Physics and Technology, School of Physics, Peking University, Beijing, 100871, China}
\address[GSI]{GSI Helmholzzentrumo f$\ddot{u}$r Schwerionenforschung, D-64291, Darmstadt, Germary}
\address[Sichuan]{Key Laboratory of Radiation Physics and Technology of the Ministry of Education, Institute of Nuclear Science and Technology, Sichuan University, Chengdu 610064, China}
\address[RCNP]{RCNP, Osaka University, Mihogaoka, Ibaraki, Osaka 567-0047, Japan}

\begin{abstract}

We report on the first measurement of the elemental fragmentation cross sections (EFCSs) of $^{29-33}\mathrm{Si}$ on a carbon target at $\sim$230~MeV/nucleon.
The experimental data covering charge changes of $\Delta Z$ = 1-4 are reproduced well   
by the isospin-dependent quantum molecular dynamics (IQMD) coupled with the evaporation GEMINI (IQMD+GEMINI) model. 
We further explore the mechanisms underlying the single-proton removal reaction in this model framework.
We conclude that the cross sections from direct proton knockout exhibit a overall weak dependence on the mass number of $\mathrm{Si}$ projectiles.
The proton evaporation induced after the projectile excitation significantly affects the cross sections for neutron-deficient $\mathrm{Si}$ isotopes,
while neutron evaporation plays a crucial role in the reactions of neutron-rich $\mathrm{Si}$ isotopes.
It is presented that the relative magnitude of one-proton and one-neutron separation energies is an essential factor that influences evaporation processes.

\end{abstract}

\end{frontmatter}

\section{Introduction}
\label{Intro}
The single-nucleon removal reaction is crucial for investigating the single-particle structure and nucleon-nucleon correlations within atomic nuclei~\cite{AUMANN2021103847}.
Compilations of experimental nucleon removal cross sections for light projectile nuclei on composite targets ($^{9}\mathrm{Be}$ or $^{12}\mathrm{C}$) at intermediate energies have revealed that the 
reduction factor $R_{\mathrm{s}}$ strongly depends on the proton-neutron Fermi-surface asymmetry, which is quantified as $\Delta S = S_{n} - S_{p}$ for neutron removal or $\Delta S = S_{p} - S_{n}$ for proton removal~\cite{PhysRevC.77.044306,PhysRevC.90.057602,PhysRevC.103.054610},
where
$S_{p}$ and $S_{n}$ are the one-proton and one-neutron separation energies of a projectile, respectively.
$R_{\mathrm{s}}$ is the ratio of experimental cross sections to theoretical predictions typically calculated using shell model spectroscopic factors and the eikonal reaction model that uses the adiabatic (or sudden) and eikonal approximations~\cite{hansen2003direct}.
However, this dependence is inconsistent with the results from lower-energy transfer reactions~\cite{PhysRevLett.104.112701, PhysRevLett.110.122503} and quasifree scattering involving ($p,2p$), ($p$,$pn$), and ($e$, $e'p$) reactions at higher energies~\cite{HOLL2019682,GOMEZRAMOS2018511,PhysRevC.70.054612}.
This inconsistency has persisted for nearly 20 years as a puzzle, attracting many attention from theoretical~\cite{PhysRevC.100.064604,PASCHALIS2020135110} and experimental studies~\cite{PANIN2019134802,DIAZCORTES2020135962,2023PohlPRL}.
For example, Hebborn~$et$~$al.$~\cite{PhysRevLett.131.212503} showed that considering theoretical optical potential uncertainties in transfer and knockout reactions yields a consistent $R_{\mathrm{s}}$-$\Delta S$ picture for loosely bound nucleons,
while Rodríguez-Sánchez~$et$~$al.$~\cite{RODRIGUEZSANCHEZ2024138559} pointed out that incorporating short-range correlations (SRCs) into dynamic intranuclear cascade models could also eliminates the $R_{\mathrm{s}}$-$\Delta S$ dependence for nucleon knockout reactions.

The inclusive single-nucleon removal cross section on a composite target at intermediate energies can involve contributions from different reaction mechanisms.
The eikonal model has well-established diffraction breakup and stripping mechanisms at incident energies down to $\sim$100~MeV/nucleon~\cite{PhysRevLett.102.232501, PhysRevC.90.064615}, but it might not accurately describe the nucleon knockout process from deeply bound states. 
Moreover, non-direct reaction processes such as multiple scattering inside of the projectile and the excitation of the residues might influence the deeply bound nucleon-removal cross section on composite target.
The eikonal model incorporates rescattering between projectile constituents only in an approximate way~\cite{PhysRevC.90.044606} and does not  
explicitly account for the evaporation channels in its calculation~\cite{PhysRevC.93.044607}.
Additionally, valence-core destruction effects caused by interactions between the residual core and removed valence nucleon in nucleon knockout reactions are not included in the eikonal model~\cite{GOMEZRAMOS2023138284}.

The single-nucleon removal reactions of $^{9}\mathrm{C}$ and $^{13}\mathrm{O}$ on $^{9}\mathrm{Be}$ target at $\sim$65~MeV/nucleon revealed that the proton evaporation components induced after projectile excitation account for roughly 17\% and 21\% of the cross sections, respectively~\cite{PhysRevC.102.044614}. 
Furthermore, in 
the $^{14}\mathrm{O}$($p$, 2$p$) reaction at $\sim$100~MeV/nucleon, the proton evaporation is evaluated to contribute nearly as much to the cross section as quasifree knockout~\cite{2023PohlPRL}.
Both experiments emphasize the significant role of proton evaporation in single-proton removal reactions.
Although in Ref.~\cite{2023PohlPRL} the proton is used as the target, the composite target is supposed to have a similar mechanism.

In this work, we aim to provide an alternative approach to understanding the mechanisms underlying single-proton removal reactions: the isospin-dependent quantum molecular dynamics (IQMD) model~\cite{HARTNACK1989303} together with the evaporation GEMINI model, $i.e.$, IQMD+GEMINI model. 
The IQMD model considers the evolution of nucleons in the mean-field potential and the nucleon-nucleon scatterings and thus includes the multiple scattering and excitation~\cite{WOLTER2022103962}.
The GEMINI models the deexcitation of the highly excited pre-fragments~\cite{CHARITY1988371}. 
Within this framework, one can assess how the projectile-like system's excitation and subsequent nucleon evaporation affect the elemental fragmentation 
cross sections (EFCSs) and how the multiple reaction mechanisms contribute to the single-proton removal cross sections.

The IQMD+GEMINI model has been extensively employed in studying heavy-ion collisions at intermediate energies, particularly in reproducing the odd-even staggering effect in spallation and fragmentation reactions~\cite{2011JunSuPRC014608, PhysRevC.97.054604}. 
To validate this model, 
we have measured the EFCSs for fragments with charge changes of $\Delta Z$ = 1-4 produced by 
$^{29-33}\mathrm{Si}$ beams on a natural carbon target at $\sim$230~MeV/nucleon. 
These measurements mark the first extension of EFCSs to the neutron-rich $\mathrm{Si}$ isotopes.
The newly obtained data, combined with the pioneering results for $^{28}\mathrm{Si}$ fragments~\cite{PhysRevC.107.024609}, 
are of significant importance for systematically exploring the patterns of fragment productions.
Encouraged by promising results with the IQMD+GEMINI model, we then address the influence of multiple reaction mechanisms in single-proton removal reactions, 
such as direct knockout, multiple scattering, and nucleon evaporation~\cite{PhysRevC.83.011601} within the model framework.

\section{Experiment}
\label{exp}
The EFCSs of $^{29-33}\mathrm{Si}$ nuclei on carbon were measured at the Heavy Ion Research Facility in Lanzhou (HIRFL)~\cite{XIA200211, ZHAN2010694c}.
A primary beam of 
320~MeV/nucleon $^{40}\mathrm{Ar}$ was impinged onto a 10 mm thick beryllium target positioned at the entrance of the Second Radioactive Ion Beam Line (RIBLL2)~\cite{SUN201878}.
The resulting cocktail beams were transported to the External Target Facility (ETF). 
Nuclei $^{29-33}\mathrm{Si}$ nuclei were identified on an event-by-event basis by a combination of magnetic rigidity ($B\rho$), energy deposition ($\Delta E$), and time of flight (TOF) measurements.
These nuclei were guided onto a natural carbon target with a thickness of 1.86 $\mathrm{g/cm^{2}}$.

A schematic layout of the experimental detector setup can be found in Ref.~\cite{PhysRevC.107.024609}. 
From the first dispersive focal plane (F1) to ETF, the TOF resolution for the incoming beam was better than 80~ps~\cite{SUN201878, Lin_2017, ZHAO201995}.
A pair of multiple sampling ionization chambers (MUSICs) was positioned upstream and downstream of the reaction target to measure $\Delta E$ of incoming and outgoing particles.
The large acceptance of MUSIC2, with an active area of 130~mm~$\times$~130~mm, guarantees the full coverage of outgoing heavy fragments.
Both detectors had $Z$ resolutions of 0.25-0.35 (FWHM) for $^{40}\mathrm{Ar}$ products.
The incident $\mathrm{Si}$ isotopes were separated using four different $B\rho$ settings, each optimized for the yield of $^{28,29}\mathrm{Si}$, $^{30}\mathrm{Si}$, $^{31,32}\mathrm{Si}$, and $^{33}\mathrm{Si}$.
Furthermore, two plastic scintillators with a central hole were positioned upstream of the target to limit the position profile of incoming
beams. 
Three multi-wire proportional chambers (MWPCs), two positioned upstream and one downstream of the reaction target, were used to determine the trajectories of incoming and outgoing particles. Positions from MWPC1 and MWPC2 were used to restrict the beam size at the target to 25~mm $\times$ 25~mm.

Measurements were also conducted without the reaction target to correct the effect of reactions in the materials other than the target (\textit{e.g.}, detectors).
The beam attenuation effect is taken into account for the precise EFCS determination for a thick target.
Following the method in Ref.~\cite{PhysRevC.107.024609}, 
the influence associated with the secondary reactions is evaluated to be negligibly small.
Taking the target thickness into account, the EFCS for an incident beam with $Z_{\mathrm{i}}$ to an element with $Z$ is calculated as
\begin{equation}
  \sigma_{\Delta Z} = \left(\frac{N_{\mathrm{F}}}{N_{0}}-\frac{N^{\mathrm{O}}_{\mathrm{F}}}{N^{\mathrm{O}}_{0}}\right)\frac{\sigma_{\mathrm{R}}}{1-\exp{(-\sigma_{\mathrm{R}}}t)}\;,
  \label{eq1}
\end{equation}
where the charge changes, $\Delta Z = Z_{\mathrm{i}}-Z$, refer to the number of protons removed from the projectile. $N_{0}$ ($N^{\mathrm{O}}_{0}$) and $N_{\mathrm{F}}$ ($N^{\mathrm{O}}_{\mathrm{F}}$) are the numbers of incident and outgoing particles for target-in (target-out) cases, respectively. 
The quantity $t$ denotes the target thickness, and is expressed in the number of target nuclei per unit area. 
$\sigma_{\mathrm{R}}$ represents the reaction cross section of incident nuclei with the target nucleus, and is 
determined by the zero-range optical-limit Glauber model (ZRGM).
The projectile's proton and neutron density distributions in this calculation are obtained from a Relativistic Hartree approach with density-dependent DD-ME2 parameter set~\cite{PhysRevC.71.024312} as inputs. 
A harmonic oscillator density distribution is employed for the $^{12}\mathrm{C}$ target.
An uncertainty of 5\% in $\sigma_{\mathrm{R}}$ results in an error of about 0.3\% in final EFCSs.

Table~\ref{tab1} summarizes the EFCSs measured in the present work for $^{29-33}\mathrm{Si}$ nuclei on carbon at $\sim$230~MeV/nucleon.
The previous results for $^{28}\mathrm{Si}$ from Ref.~\cite{PhysRevC.107.024609} are also included.
For the EFCSs with $\Delta Z$ = 1-4, the statistical errors range from 4.9\% to 11.4\%.
The systematic errors associated with the peak decoupling method are evaluated to be below 8.2\%.
Moreover, the secondary reactions in the target contribute to the systematic errors by about 2.5\%, which are 
smaller than the statistical errors.
The total errors of EFCSs are estimated by adding the statistical and systematic errors in quadrature.
\begin{table}[!htpb]\small
    \centering
    \caption{Summary of elemental fragmentation 
     cross sections (EFCSs) for $^{29-33}\mathrm{Si}$ nuclei on carbon at $\sim$230~MeV/nucleon. The literature data taken from Ref.~\cite{PhysRevC.107.024609} for $^{28}\mathrm{Si}$ are included. The error includes both statistical and systematic errors.}
    \begin{threeparttable}[b]
    \setlength{\tabcolsep}{1.6mm}{
    \begin{tabular}{cccccc}
    \hline
    \hline
\multirow{2}*{isotopes}
        &Incident energy
        &$\sigma_{\Delta Z=1}$
        &$\sigma_{\Delta Z=2}$     
        &$\sigma_{\Delta Z=3}$
        &$\sigma_{\Delta Z=4}$\\
        & MeV/nucleon & mb & mb & mb & mb\\
    \hline
\hline
$^{33}$Si & 235 &189$\pm$17    &131$\pm$16  &85$\pm$10  & 96$\pm$10\tnote{1} \\
$^{32}$Si & 232 &208$\pm$11    &128$\pm$9   &87$\pm$7   & 87$\pm$7\tnote{1} \\
$^{31}$Si & 229 &158$\pm$13    &125$\pm$11  &92$\pm$10   & 79$\pm$8\tnote{1} \\
$^{30}$Si & 225 &170$\pm$11    &142$\pm$10   &83$\pm$8   & 93$\pm$7\tnote{1} \\
$^{29}$Si & 217 &161$\pm$11    &144$\pm$13  &81$\pm$11   & 89$\pm$8\tnote{1} \\
$^{28}$Si & 218 &140$\pm$8     &155$\pm$9   &91$\pm$7   & 101$\pm$9\tnote{2} \\
    \hline
    \hline
    \end{tabular}
    \label{tab1}
   \begin{tablenotes}
     \item[1] Present work.
     \item[2] From Ref.~\cite{PhysRevC.107.024609}.
   \end{tablenotes}}
  \end{threeparttable}
\end{table}

\section{Theoretical approach}
\label{approach}
In the IQMD model, the many-body state is represented by a simple product wave function of single-particle states in a fixed Gaussian shape without anti-symmetrization
\begin{equation}
    \centering
    \begin{split}
    &\psi\left(\textbf{r},t \right) = \prod_{i=1}\phi_{i}\left(\textbf{r},t\right)\;, \\
    &\phi_{i}(\textbf{r},t) = \frac{1}{\left(2\pi L\right)^{3/4}}\exp\left\lbrack-\frac{\lbrack \textbf{r}-\textbf{r}_{i}(t)\rbrack^{2}}{4L}\right\rbrack\exp\left[\frac{i\textbf{r}\cdot\textbf{p}_{i}(t)}{\hbar}\right]\;,
    \end{split}
    \label{eq2}
\end{equation}
where $\textbf{r}_{i}$ and $\textbf{p}_{i}$ represent the average values of the positions and momenta of the $i$th nucleon.
$L$ is associated with the extension of the wave packet.
The Wigner transform of the wave function derives the 1-body Wigner function.
Then the density distribution function $\rho$ of the system, which will be applied in the potential energy density functional, can be expressed as
\begin{equation}
    \centering
    \rho(\boldsymbol{\mathrm{r}}) = \sum_{i}\frac{1}{(2\pi L)^{3/2}}\exp{\left\lbrack -\frac{\lbrack\boldsymbol{\mathrm{r}}-\boldsymbol{\mathrm{r}}_{i}(t)\rbrack^{2}}{2L}\right\rbrack}\;.
    \label{eq3}
\end{equation}

The Gaussian wave packets yield the following equations of motion derived from the time-dependent variational principle
\begin{equation}
\begin{split}
&\dot{\textbf{r}}_{i} = \nabla_{\textbf{p}_{i}} H , \\
&\dot{\textbf{p}}_{i} = -\nabla_{\textbf{r}_{i}} H .\\
\end{split}
\label{eq4}
\end{equation}
Here, $H$ represents the total $N$-body Hamiltonian, which consists of the kinetic energy, Coulomb potential energy, and local nuclear potential energy.
The nuclear potential energy is a Skyrme-type integration of the potential energy density functional.
The parameters are derived from the Skyrme parameters Sly6~\cite{CHABANAT1997710}
and provide a compressibility of 271~MeV at saturation density for isospin-symmetric nuclear matter. 
They are the same as those utilized in Ref.~\cite{PhysRevC.107.024609}.

In addition to the mean-field propagation, the nucleon-nucleon scattering and the Pauli blocking effect of the final state are incorporated.
The nucleon-nucleon scattering drives the stochastic evaluation and leads to enhanced fluctuations of the one-body density.
Thus, in the philosophy of the IQMD model, one goes beyond the mean-field approach and includes correlations and fluctuations.
The procedure of nucleon-nucleon scattering involves two steps: first, to determine if two test particles collide in a given time step, and second, to check whether the final state of the collision is allowed by the Pauli principle.
Two nucleons collide in a given time step if they reach a
distance of the closest approach given by the cross section 
within that step. The choice of the final momenta of the nucleons is stochastic, with the condition that the total energy and momentum are conserved.

The calculation is a two-step process, including the dynamic evolution described by the IQMD and statistical decay by the GEMINI.
The dynamic evolution time is selected as 100~$\mathrm{fm/c}$.
The GEMINI code receives inputs of atomic number, mass number, and excitation energy for each pre-fragment outputted by the IQMD model.
If the excitation energy exceeds zero, the pre-fragment will decay by evaporating light particles  ($\textit{e.g.}$, $n$, $p$, $\alpha$) and/or emitting $\gamma$ rays.
The particle decay widths are modeled using the Hauser-Feshbach formalism~\cite{PhysRev.87.366} for evaporation.
For emitting a light particle ($Z_{1},A_{1}$) with a spin of $J_{1}$ from a system ($Z_{0},A_{0}$) with an excitation energy $E^{*}$ and spin $J_{0}$, resulting in the residual system ($Z_{2},A_{2}$) with spin $J_{2}$, the decay width $\Gamma_{J_{2}}$ is given by
\begin{equation}
    \begin{aligned}
    \Gamma_{J_{2}}(Z_{1},A_{1},Z_{2},A_{2}) & = \frac{2J_{1}+1}{2\pi\rho_{0}}\sum^{J_{0}+J_{2}}_{l=|J_{0}-J_{2}|}\int^{E^{*}-B-E_{\mathrm{rot}}}_{0}  \\
    & \times T_{l}(\varepsilon)\rho_{2}(E^{*}-B-E_{\mathrm{rot}}-\varepsilon,J_{2})d\varepsilon\;,
    \label{eq5}
    \end{aligned}
\end{equation}
where $l$ and $\varepsilon$ represent the orbital angular momentum and kinetic energy of the emitted particle, respectively.
$E_{\mathrm{rot}}$ accounts for the rotational energy of the residual system. 
$\rho_{0}$ and $\rho_{2}$ denote the level densities of the initial and residual systems, respectively.
The level densities follow the Fermi-gas prescription.~\cite{DILG1973269}.
The binding energy $B$ is calculated from the nuclear masses.
Nuclear masses and level densities with pairing correlations are adopted in the GEMINI model~\cite{2011JunSuPRC014608}. 
The tabulated masses are applied~\cite{AUDI2003337}.
Detailed descriptions of the GEMINI model can be found in Ref.~\cite{CHARITY1988371}.

\section{Results and discussion}
\label{dis}

\begin{figure}[!htpb]
    \centering
    \includegraphics[width=0.40\textwidth, angle=-0]{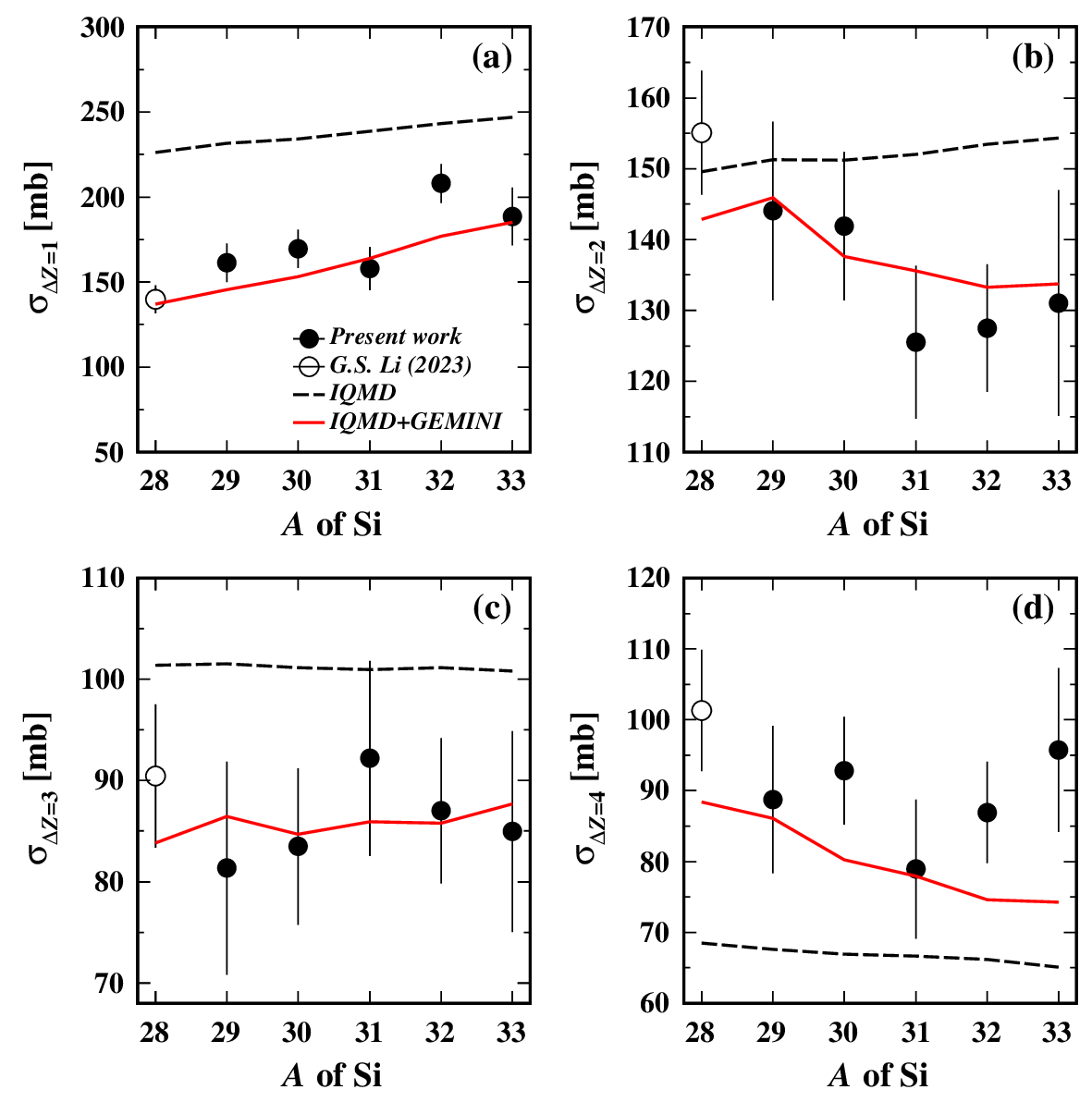}
    \caption{Elemental fragmentation cross sections (EFCSs) for fragments with $\Delta Z$ = 1-4, $i.e.$, (a) $\mathrm{Al}$, (b) $\mathrm{Mg}$, (c) $\mathrm{Na}$, and (d) $\mathrm{Ne}$ isotopes, as a function of mass number $A$ of $^{28-33}\mathrm{Si}$ on carbon at $\sim$230~MeV/nucleon. The present results are shown by filled circles, while the literature data taken from Ref.~\cite{PhysRevC.107.024609} are shown by open symbols. For comparison, the IQMD and the IQMD+GEMINI model predictions are shown as dashed and solid lines, respectively.}
    \label{fig1}
\end{figure}

Figures~\ref{fig1}(a)--\ref{fig1}(d) present the EFCSs with $\Delta Z$ = 1-4 fragments from $^{28-33}\mathrm{Si}$ as a function of projectile mass number $A$. 
The experimental data from the present study are represented by black-filled circles, while the data from Ref.~\cite{PhysRevC.107.024609} are shown by open symbols.
For $\Delta Z$ = 1 residues, the EFCSs tend to increase as the mass number increases.
Similar trends of EFCSs at $\Delta Z$ = 1 are observed for $\mathrm{Ca}$ and $\mathrm{Ti}$ isotopes colliding with carbon at $\sim$300~MeV/nucleon~\cite{YAMAKI2013774},
as well as boron fragments produced from $^{12-16}\mathrm{C}$ on carbon at $\sim$240~MeV/nucleon~\cite{Jin2022CPC}.
Concerning the data at $\Delta Z$ = 2-4, it is currently challenging to establish a definitive trend due to the relatively large uncertainties.

In Figs.~\ref{fig1}(a)--\ref{fig1}(d), the results of IQMD and IQMD+GEMINI model calculations are shown by the black dotted lines and by the red solid lines.
One sees that the IQMD results exhibit a monotonically changing trend as $\Delta Z$ varies from 1 to 4 but fail to reproduce the experimental results.
In contrast, the IQMD+GEMINI results reproduce the experimental data pretty well.
This indicates the necessity of the statistical decay process beyond
the dynamics in heavy-ion collisions~\cite{PhysRevC.98.014610, PhysRevC.105.024608}.
We also assessed predictions from the empirical parametrizations such as
EPAX3~\cite{PhysRevC.86.014601}, modified EPAX2~\cite{ZHANG201359}, FRACS~\cite{PhysRevC.95.034608}, as well as the statistical models  ABRABLA07~\cite{GAIMARD1991709} and NUCFRG2~\cite{WILSON199495},
but none of them predict the EFCSs as effective as the IQMD+GEMINI model. 
Moreover, we find that the IQMD+GEMINI model can describe the other cross sections for different projectile-target combinations at various incident energies
~\cite{IANCU2005525,PhysRevC.77.034605,PhysRevC.56.388,PhysRevC.107.024609,SAWAHATA2017142,ZEITLIN2007341}.
Similar evidences of the significance of charged-particle evaporation were reported in the study of charge-changing reactions~\cite{PhysRevC.106.014617,ZHAO2023138269,ZHANG20241647}.

To study the dynamics leading to the projectile-like system excitation and sequential nucleon evaporation on EFCSs,
it is important to note the variation in the relationship between the IQMD and IQMD+GEMINI results.
As seen in Figs.~\ref{fig1}(a)-\ref{fig1}(c), the IQMD+GEMINI model consistently predicts smaller cross sections for $\Delta Z$ = 1-3 compared to those obtained from the IQMD model.
The discrepancies between them decrease for larger $\Delta Z$. 
Then, in contrast, the IQMD+GEMINI cross sections are larger than those of IQMD for $\Delta Z$ = 4, as seen in Fig.~\ref{fig1}(d).
These observations can be explained by the higher excitation energy allowing for multiple sequential decay processes.
Knocking out more protons will lead to more possible decay pathways.
The decay path ending as a final fragment with higher stability will be chosen with a higher probability.
Let the pre-fragment cross section to an element with $Z$ in the IQMD model be $\sigma^{\mathrm{Pre}}_{Z}$, then the fragment cross section $\sigma^{\mathrm{Fin}}_{Z}$ in the IQMD+GEMINI model
can be written as $\sigma^{\mathrm{Fin}}_{Z}$ = $\sigma^{\mathrm{Pre}}_{Z}$
- $\sigma^{\mathrm{evap}}_{<Z}$ + $\sigma^{\mathrm{evap}}_{>Z}$.
Here $\sigma^{\mathrm{evap}}_{<Z}$ represents the cross section that goes to lower-$Z$ final fragments from pre-fragments with $Z$ through the evaporation processes, which consequently reduces $\sigma^{\mathrm{Pre}}_{Z}$.
$\sigma^{\mathrm{evap}}_{>Z}$ denotes the cross section from higher-$Z$ pre-fragments that evaporate continuously to produce final residues with $Z$,
leading to an increase of $\sigma^{\mathrm{Pre}}_{Z}$.
In Figs.~\ref{fig1}(a)-\ref{fig1}(d), 
$\sigma^{\mathrm{Pre}}_{Z}$ decreases rapidly with increasing $\Delta Z$.
This results in the reduction of $\sigma^{\mathrm{evap}}_{<Z}$ since it constitutes a portion of $\sigma^{\mathrm{Pre}}_{Z}$.
In contrast, $\sigma^{\mathrm{evap}}_{>Z}$ increases with $\Delta Z$ due to the increase in the number of pre-fragments.
Consequently, the diversity of the decay pathways causes $\sigma^{\mathrm{Fin}}_{Z}$ to gradually approach $\sigma^{\mathrm{Pre}}_{Z}$ for 
$\Delta Z$ = 1-3, but to surpass $\sigma^{\mathrm{Pre}}_{Z}$ for $\Delta Z$ = 4.  
The shift in the relationship between the IQMD and the IQMD+GEMINI model calculations indicates the significance of the evaporation processes.

\section{Single-proton removal in the IQMD+GEMINI model}
\label{Model_Pred}
The good agreement between the IQMD+GEMINI results and the experimental data allows for a detailed study of the underlying mechanisms of nucleon removal reactions. 
The IQMD+GEMINI model categorizes fragmentation into two distinct processes: hard nucleon collision and evaporation, which are treated by IQMD and GEMINI models, respectively. Under this framework, 
the emission of nucleon(s) together with the final residue can occur over different timescales. 
For the sake of clarity, 
we target the scenarios where only one proton is ejected from the $^{28}\mathrm{Si}$ projectile as an example.
Thus the reaction mechanisms for single-proton removal from $^{28}\mathrm{Si}$ in collision with carbon to produce the final fragment $^{27}\mathrm{Al}$ 
can be classified into three processes:
\begin{itemize}
    \item [(1)]
    A proton is knocked out from initial $^{28}\mathrm{Si}$ by a hard collision, leaving an excited pre-fragment $^{27}\mathrm{Al}^{*}$.
    The pre-fragment $^{27}\mathrm{Al}^{*}$ then decays via possible $\gamma$-ray emission and ends in the final residue $^{27}\mathrm{Al}$. This process is identified as the direct ``proton knockout".
    \item [(2)] The projectile-target collisions can also produce a pre-fragment having the identical mass and atomic number as initial $^{28}\mathrm{Si}$ but in an excited state ($i.e.$, $^{28}\mathrm{Si}^{*}$). 
    Excited $^{28}\mathrm{Si}^{*}$ then de-excites by evaporating a proton and possible $\gamma$-rays with $^{27}\mathrm{Al}$ being the final fragment.
    This process is referred to as the ``proton evaporation".
    The strength of the evaporation is strongly correlated with the one-proton separation energy $S_{p}$.
    \item [(3)] The other processes, such as the ``charge-exchange", also contribute to the single-proton removal reaction.
    During the ($p$, $n$) charge-exchange, the excited $^{28}\mathrm{Al}^{*}$ is initially produced in the hard collision stage and then de-excites by evaporating a neutron and possible $\gamma$-ray, resulting in the formation of $^{27}\mathrm{Al}$ final fragment.
\end{itemize}

\begin{figure}[!htpb]
    \centering
    \includegraphics[width=0.425\textwidth, angle=-0]{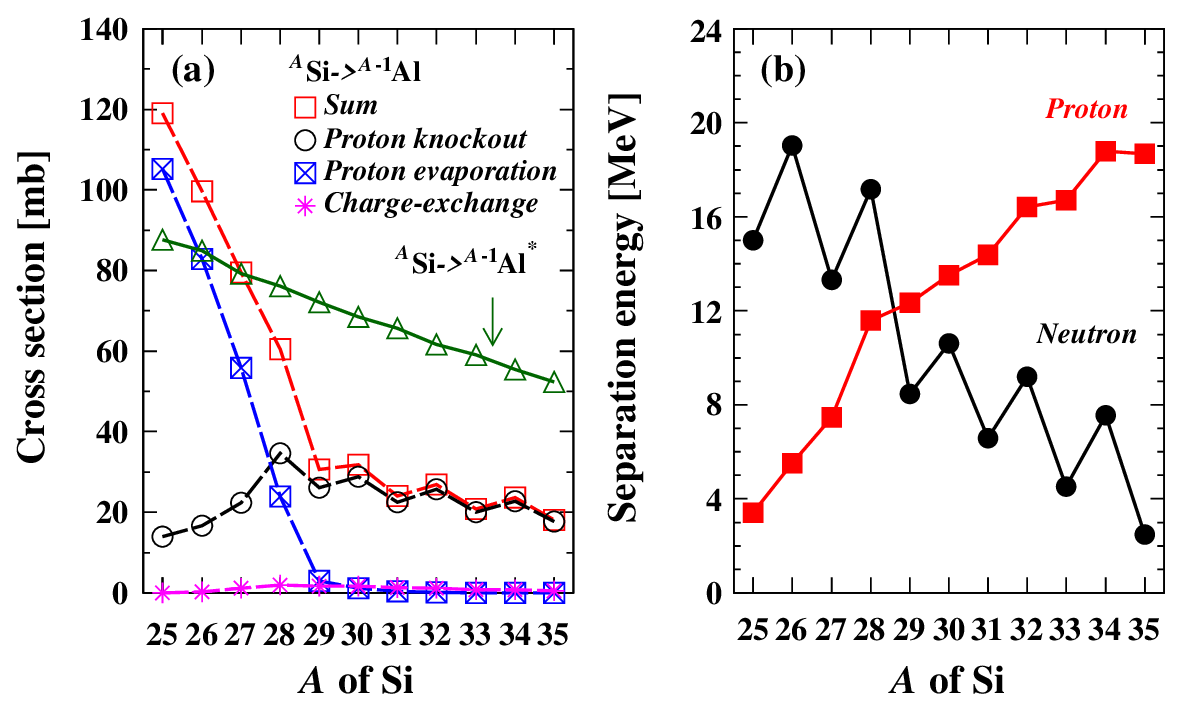}
    \caption{(a) The cross sections for proton knockout, proton evaporation, and charge-exchange processes, and their summation for $^{A-1}\mathrm{Al}$ final fragments as a function of mass number $A$ of $\mathrm{Si}$ ($A$ = 25-35). For comparison, the cross sections of pre-fragments $^{A-1}\mathrm{Al}^{*}$  are also shown by the open green triangles. 
    (b) One-proton and one-neutron separation energies ($S_{p}$ and $S_{n}$) of $^{25-35}\mathrm{Si}$ nuclei are presented.
    Proton(neutron) evaporation predominantly influences single-proton removal reactions when $S_{p}$($S_{n}$) is less than $S_{n}$($S_{p}$). For the details, see the text.}
    \label{fig2}
\end{figure}

To evaluate the influences of those reaction mechanisms,
we calculate the
cross sections for proton knockout, proton evaporation, and charge-exchange processes that produce $^{A-1}\mathrm{Al}$ fragments from the reactions of $^{A}\mathrm{Si}$ ($A$ = 25-35) with a carbon target at 230 MeV/nucleon using the IQMD+GEMINI model.
As illustrated in Fig.~\ref{fig2}(a),
the charge-exchange cross sections typically amount to less than 6\% of the total cross sections.
The cross sections from the proton knockout depend weakly on the projectile mass number.   
This pattern significantly differs from the trend observed in the proton evaporation,
where the cross sections decrease monotonically as the mass number increases up to $A$ = 29 and then 
remain at very low values. 
This highlights the significant influence of the proton evaporation on single-proton removal reactions for the neutron-deficient $\mathrm{Si}$ isotopes.

The trends in cross sections can be linked to the nucleon separation energies. Figure~\ref{fig2}(b) presents the one-proton and one-neutron separation energies ($S_{p}$ and $S_{n}$) of $^{25-35}\mathrm{Si}$ nuclei. The $S_{p}$ gradually increases with $A$, while $S_{n}$ behaves in the opposite manner. Moreover, $S_{p}$ is systematically smaller than $S_{n}$ for $A \leq 28$, with the discrepancies being more pronounced toward the neutron-deficient side.
This is exactly in accordance with the significant increase in cross sections for proton evaporation.
Consequently, the projectile excitation followed by proton evaporation predominantly contributes to single-proton removal cross sections for $A \leq$ 28.

As for $A\geq$ 29, where $S_{p}$ is consistently larger than $S_{n}$, the proton knockout plays a significant role in determining the cross sections rather than the proton evaporation. 
However, the proton knockout contributes much smaller cross sections for $^{A-1}\mathrm{Al}$ fragments compared to those
expected from the pre-fragments $^{A-1}\mathrm{Al}^{*}$ generated by hard collisions, as shown in Fig.~\ref{fig2}(a).
The fragments $^{A-1}\mathrm{Al}$ cross sections change almost in parallel with those of pre-fragments $^{A-1}\mathrm{Al}^{*}$ for $A\geq$ 29, where 
the contribution from the proton evaporation is negligibly small.
About 60\% of pre-fragments $^{A-1}\mathrm{Al}^{*}$ cross sections are lost by evaporation.
Considering the smaller $S_{n}$ (or larger $S_{p}$) for the neutron-rich $\mathrm{Si}$ isotopes,
it is believed that the neutron(s) evaporation from pre-fragments dominates and strongly reduces the pre-fragment cross sections.
Moreover, the proton knockout cross sections typically decrease toward the neutron-deficient $\mathrm{Si}$ isotopes with $A \leq$ 28.
This is attributed to the increased contribution of proton evaporation
from $^{A-1}\mathrm{Al}^{*}$ pre-fragments to form lower-($A-1$) final fragments,
similar to the increase in proton evaporation cross sections from $\mathrm{Si}$ pre-fragments.
The proton(neutron) evaporation becomes significant when the $S_{p}$($S_{n}$)
is smaller than $S_{n}$($S_{p}$).

\begin{figure}[!htpb]
    \centering
    \includegraphics[width=0.40\textwidth, angle=-0]{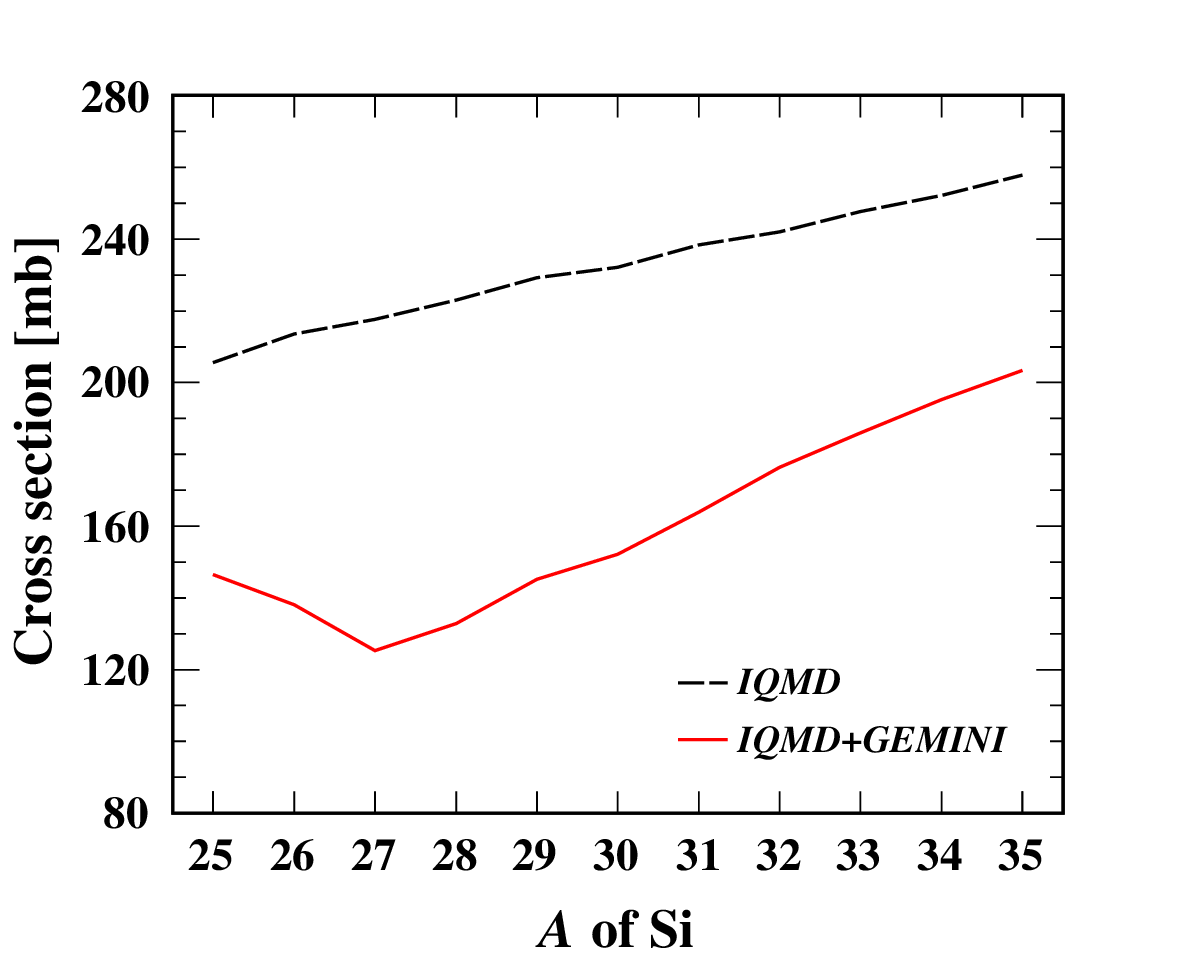}
    \caption{The EFCS results of $\Delta Z$ = 1 pre-fragments (black line) and final fragments (red line) produced by $^{25-35}\mathrm{Si}$ nuclei on carbon at 230~MeV/nucleon, which are simulated by using the IQMD and IQMD+GEMINI models, respectively.}
    \label{fig3}
\end{figure}

In Fig.~\ref{fig3}, the EFCS results from the IQMD model for $\Delta Z$ = 1 pre-fragments produced by $^{25-33}\mathrm{Si}$ on carbon at 230~MeV/nucleon tend to increase as mass number $A$ increases.
In contrast, the IQMD+GEMINI results exhibit a distinct kink at $A$ = 27.
This suggests that the 
proton evaporation predominantly influences fragment cross sections when $A$ is less than 27,
whereas neutron evaporation becomes increasingly significant for $A$ greater than 27.

\begin{figure}[!htpb]\small
    \centering
    \includegraphics[width=0.40\textwidth, angle=-0]{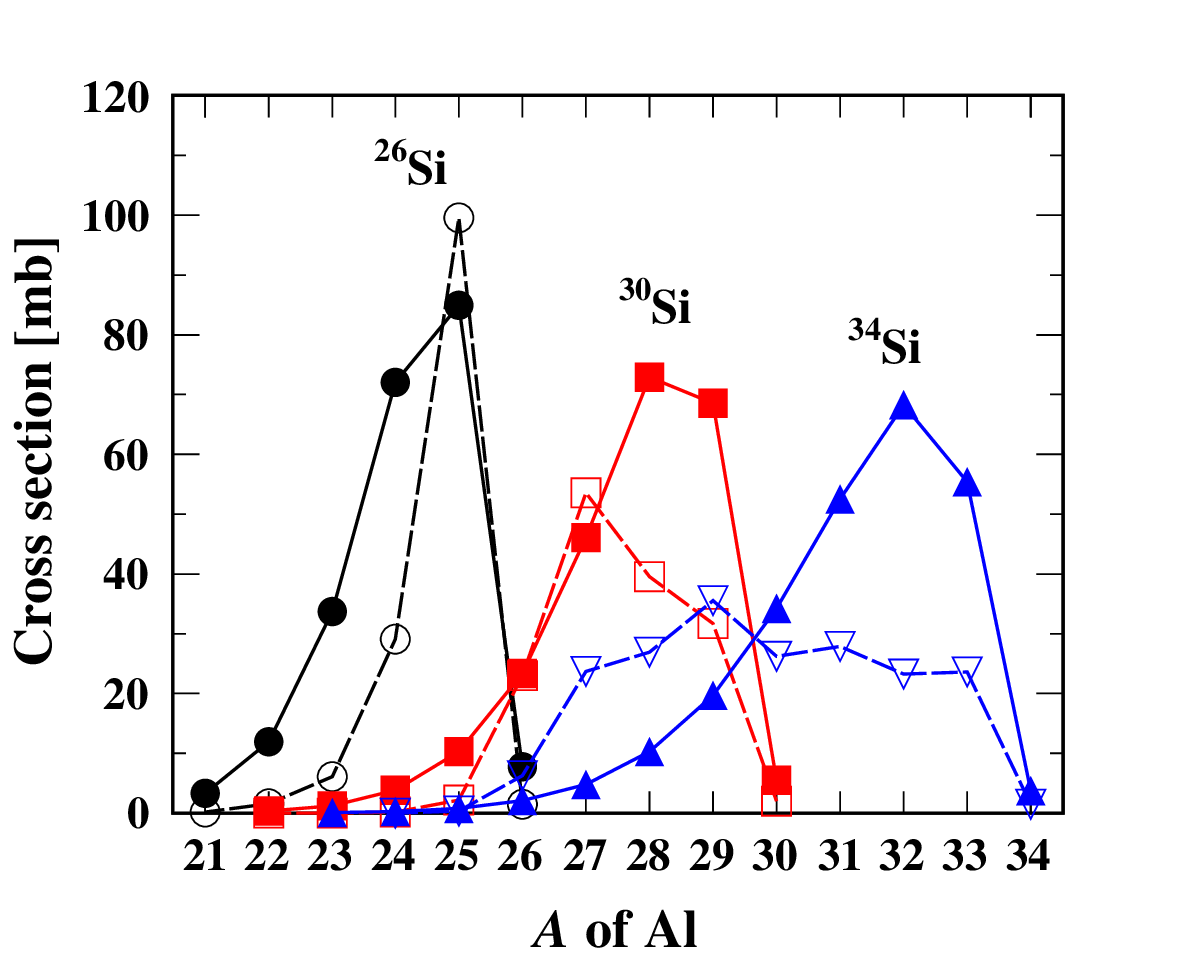}
    \caption{Fragment (open symbol) and pre-fragment (filled symbol) cross sections of $\mathrm{Al}$ isotopes produced by $^{26}\mathrm{Si}$, $^{30}\mathrm{Si}$ and $^{34}\mathrm{Si}$ projectiles colliding with carbon at 230~MeV/nucleon. The results from the fragmentation of these three projectiles are represented by black, red, and blue data points, respectively.}
    \label{fig4}
\end{figure}

The aforementioned discussion in Fig.~\ref{fig2}(a) is supported by examining the fragment cross sections of different isotopes.
Figure~\ref{fig4}
presents such cross sections for $^{26}\mathrm{Si}$, $^{30}\mathrm{Si}$ and $^{34}\mathrm{Si}$ projectiles as examples.
Both fragment (open symbols) and pre-fragment (filled symbols) cross sections are shown for comparison.
In the fragmentation of $^{26}\mathrm{Si}$ projectile, the cross section of fragment $^{25}\mathrm{Al}$  
is larger than that of pre-fragment $^{25}\mathrm{Al}^{*}$
This indicates that the proton evaporation from excited $^{26}\mathrm{Si}^{*}$ dominates the fragment cross section.
The cross section of fragment $^{24}\mathrm{Al}$ is much smaller than that of pre-fragment $^{24}\mathrm{Al}^{*}$
and thus shows that the cross section is reduced by the evaporation process but is not much increased from neutron evaporation of heavier $\mathrm{Al}$ isotopes.
A clear difference is seen in $^{34}\mathrm{Si}$ projectile where fragment cross sections of $\mathrm{Al}$ isotopes with neutron removal are as large as those without neutron removal.
In particular, the cross sections of fragments $\mathrm{Al}$ with mass number less than 29 are larger than those of pre-fragments, indicating the significant contribution from the neutron evaporation of heavier $\mathrm{Al}$ pre-fragments.
Fragmentation of $^{30}\mathrm{Si}$ clearly shows the transition from the neutron-deficient to the neutron-rich side of $\mathrm{Al}$ isotopes.  
Hence, neutron evaporation significantly influences the fragmentation of neutron-rich nuclei,
whereas proton evaporation dominates the fragmentation of neutron-deficient nuclei.

\section{Summary}
\label{summary}
We have presented the ETCSs of fragments with $\Delta Z$ = 1-4 produced by $^{29-33}\mathrm{Si}$ on a carbon target at $\sim$230~MeV/nucleon. 
These measurements mark the first extension of EFCSs to the neutron-rich $\mathrm{Si}$ isotopes. 
The IQMD+GEMINI model reproduces the experimental data pretty well.
Therefore, we explored the dynamics, the production of pre-fragments, and the sequential nucleon evaporation by using the IQMD+GEMINI model.
In particular, we studied the reaction mechanisms in single-proton removal reactions.
We concluded that neutron evaporation from the pre-fragment strongly reduces the cross sections of neutron-rich nuclei,
while 
proton evaporation from pre-fragment increases the cross sections of neutron-deficient nuclei.
Currently, there is no experimental data available for neutron-deficient $\mathrm{Si}$ incident fragment cross sections. Such data would provide a crucial insights into our understanding of the evaporation processes.
We showed that the IQMD+GEMINI model offers a new view of underlying mechanisms in single-proton removal reactions and other fragment reactions.

\section*{\uppercase{Acknowledgments}}
\label{acknowledgments}
The authors express their gratitude to the staff of the HIRFL-CSR accelerator for their dedicated efforts and assistance in maintaining a stable beam condition throughout the experiment.
The authors are grateful to T. Yamaguchi for generously providing the experimental data in Ref.~\cite{YAMAKI2013774}.
Special thanks are also due to Z. Z. Li and Y. F. Niu for their valuable contributions and assistance in determining the nucleon density distributions.
This work 
was supported partially by the National Natural Science Foundation of China (Nos. 12325506, 11961141004, 
11922501, 11475014, 11905260, the National Key R\&D program of
China (No. 2016YFA0400504).
X. D. Xu acknowledges the support of the Western Light Project of the Chinese Academy of Sciences.

\bibliographystyle{elsarticle-num}

\end{document}